\documentclass[10pt,a4paper,english,twocolumn]{IEEEtran} 
\usepackage[T1]{fontenc}
\usepackage{color, colortbl}
\usepackage{babel}
\usepackage{verbatim}
\usepackage{textcomp}
\usepackage{amsmath}
\usepackage{amssymb}
\usepackage{amstext}
\usepackage{graphicx}
\usepackage{tabulary}
\usepackage{caption}

\definecolor{Gray}{gray}{0.95}
\definecolor{LightCyan}{rgb}{0.8,0.85,1}
\definecolor{LightBlue}{rgb}{0.6,0.6,1}
\usepackage[font=small]{caption}
\usepackage{enumitem}

\makeatletter

\pdfpageheight\paperheight
\pdfpagewidth\paperwidth

\usepackage{epstopdf}
\usepackage{cite}
\usepackage{citesort}
\usepackage[normalem]{ulem}
\usepackage{balance}
\addto\captionsenglish{\renewcommand{\figurename}{Fig.}}

\makeatother

\begin{document}
\title{IEEE 802.11be: Wi-Fi 7 Strikes Back}

\author{
{Adrian Garcia-Rodriguez},
{David L$\acute{\textrm{o}}$pez-P$\acute{\textrm{e}}$rez}, 
{Lorenzo Galati-Giordano}, and
{Giovanni Geraci}
\thanks{{A. Garcia-Rodriguez} and {D. L$\acute{\textrm{o}}$pez-P$\acute{\textrm{e}}$rez} were with Nokia Bell Labs, Dublin, Ireland, when the work was carried out.}
\thanks{{L. Galati-Giordano} is with Nokia Bell Labs, Dublin, Ireland.}
\thanks{{G. Geraci} is with Universitat Pompeu Fabra, Barcelona, Spain.}
\thanks{This work was partly supported by the H2020-MSCA-ITN-2018 Grant 812991, by MINECO Project RTI2018-101040-A-I00, and by the Junior Leader Fellowship Program from ``la Caixa'' Banking Foundation.}
}

\maketitle

\begin{abstract}
As hordes of data-hungry devices challenge its current capabilities, Wi-Fi \emph{strikes back} with 802.11be, alias Wi-Fi~7. This brand-new amendment promises a (r)evolution of unlicensed wireless connectivity as we know it. With its standardisation process being consolidated, we provide an updated digest of 802.11be essential features, vouching for multi-AP coordination as a must-have for critical and latency-sensitive applications. We then get down to the nitty-gritty of one of its most enticing implementations---coordinated beamforming---, for which our standard-compliant simulations confirm near-tenfold reductions in worst-case delays. 
\end{abstract}


\section{Introduction\label{sec:Introduction}}

Back in 1943, psychologist Abraham Maslow published a study on the hierarchy of human needs, stating that before achieving full use of one's talents and interests, four needs must first be met \cite{Maslow1943}. 
His theory is best illustrated by a pyramid, listing from the base: physiological needs, safety, belonging, and esteem. 
Nowadays, one could provocatively add one more layer at the bottom of Maslow's pyramid: Wi-Fi. 
Beyond food, shelter, and clean water, access to wireless connectivity is regarded as a must in our globalised society. 
While we would argue that humans do not need Internet more than breathable air, the importance of Wi-Fi is unquestionable. During forced confinement, many of us resorted to Wi-Fi to be in touch with our loved ones, to place online orders that kept small businesses afloat, and to keep fit by taking live-streaming yoga classes. After all, this article could hardly have been written without Wi-Fi, and as you read it, chances are you are using Wi-Fi too.

Relied on by billions of people every day, Wi-Fi carries most of the global data traffic in an ever-expanding variety of applications. There will be nearly 628 million public Wi-Fi hotspots by 2023, one out of ten equipped with Wi-Fi's sixth version based on the IEEE 802.11ax specification \cite{Cisco2020,2018IeeeStandardAxDraft}. 
As the popularity and capabilities of Wi-Fi grow, so will the demand for wireless services. More and more households will accommodate smart-home appliances besides 8K displays and virtual reality gadgets, turning into dense environments with many concurrently connecting devices. Enterprises will be sharply increasing the amount of data collected across their premises, targeting enhanced manufacturing processes and augmented productivity. Importantly, such cross-factory-floor communications will afford very low latency to enable machinery synchronisation and real-time control \cite{WBA2019}. Live video will take up a large share of the global IP traffic, with high-quality conferencing enabling more remote-friendly work, education, and healthcare in a post-pandemic world.

Our call for gigabits per second---delivered reliably to every nook and cranny of our apartments and enterprise spaces---is motivating the development of a new Wi-Fi 7 generation based on
IEEE 802.11be Extremely High Throughput (EHT). Since the introduction of 802.11be to our research community \cite{LopGarGal2019,KhoLevAky2020}, much has been cooking in the regulation, certification, and standardisation bodies. The United States Federal Communications Commission (FCC) has made new spectrum available in the 6~GHz frequencies for unlicensed use, and the Wi-Fi Alliance---a worldwide network of companies driving the adoption and evolution of Wi-Fi through a certified seal---is soon expected to provide worldwide interoperability certification for Wi-Fi~6 devices to function in such a new band. Meanwhile, key experts are gathering at virtual IEEE meetings to define the building blocks of the 802.11be standard. 


In this article, we forecast how the Wi-Fi of the future will be, starting with an update on the status quo. We then review in detail the latest concrete decisions on the technical features to be adopted in the 802.11be amendment, along with the new estimated timeline for their development. We also discuss one of the most appealing enablers to complement improved peak throughput with boosted network efficiency, lower latency, and higher reliability: multi-access point (AP) coordinated beamforming (CBF). In particular, we shed light on the details of its potential implementation, and we share standard-compliant simulation results that quantify the latency gains it attains in a realistic digital enterprise setup.
%




 \begin{figure*}[t!]
\begin{centering}
	\includegraphics[width=1.9\columnwidth]{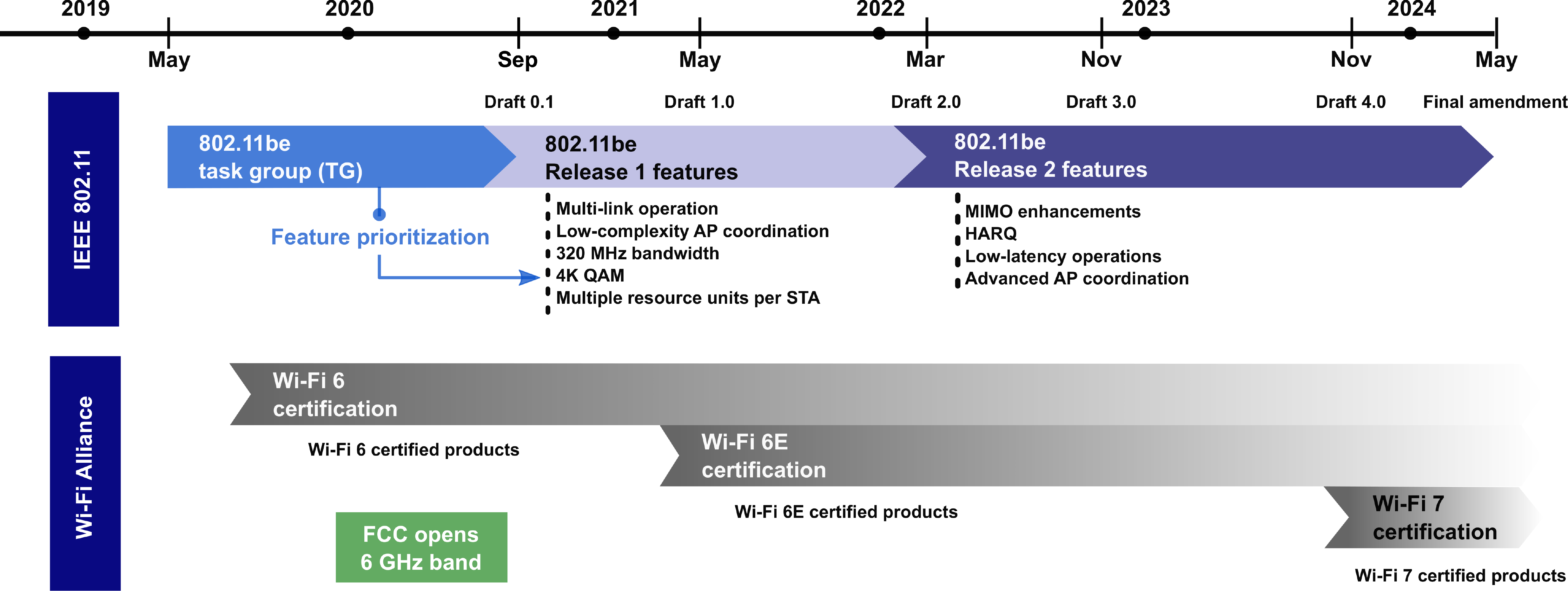} 
	\caption{Illustration of the current Wi-Fi 7 standardisation, certification, and commercialisation timelines.}
\label{fig:timeline}
\end{centering}
\end{figure*}

\section{A Concise Update on Wi-Fi \label{sec:updates}}

Dealing with tighter requirements in high-density scenarios is one of the most ambitious goals Wi-Fi must meet to maintain its position as a top trending wireless technology. State-of-the-art Wi-Fi~6, based on IEEE 802.11ax, faces crowded deployments by improving network efficiency and battery consumption through features, among others, such as Orthogonal Frequency Division Multiple Access (OFDMA) and uplink---as well as downlink---multi-user MIMO \cite{2018IeeeStandardAxDraft}. 
With 802.11ax yet to achieve its final ratification, delayed till late 2020, Wi-Fi's stakeholders are already eyeing a further two-step improvement over Wi-Fi~6. The first one will be brought by Wi-Fi~6E, as governments worldwide open up new frequencies for unlicensed use. The second one will be the new 802.11be amendment, likely to be certified as Wi-Fi~7.

\subsection{Wi-Fi~6E: Our Traffic Gets a New Lane}

For over 20 years, Wi-Fi has been working by broadcasting airwaves over two bands: 2.4 and 5~GHz. 
In April 2020, the FCC unanimously cleared the way for a third band: 5.925--7.125~GHz \cite{FCC2020}. 
Such added spectrum, referred to as the 6~GHz band, nearly quadruples the amount of available bandwidth. In addition to the higher number of available channels, a key difference of the newly opened frequencies lies in their shorter propagation range, which may be particularly suitable to provide Basic Service Set (BSS) isolation in dense and challenging environments like transportation hubs, sports arenas, and business complexes. The new 6~GHz band will be accessed by unlicensed devices under rules devised to protect incumbent services. Among them, there will be a mandatory contention-based protocol for outdoor use, and limits, not only on the total emitted power, but also on its power spectral density to discourage inefficient use of narrow channels.

While the FCC decision may have given the United States an initial lead on the 6~GHz market, other regions including Europe and Asia-Pacific are also exploring unlicensed access to this band. In the meantime, Wi-Fi~6 is ready to utilise the 6~GHz spectrum as it becomes available worldwide, and devices equipped with the chips and radios needed to operate in the new band will get a `6E' seal, the E standing for extension. The Wi-Fi Alliance plans to launch its Wi-Fi~6E certification in early 2021, with more than 300 million compliant devices expected to come to market the same year. 
Remarkably, since only 6E devices will be initially able to operate in the 6~GHz band, they will avail---at least at the beginning---of a pristine, low-interference setting.

\subsection{Wi-Fi~7: (Not Just) Extremely High Throughput}

Despite its name, 802.11be Extremely High Throughput will be much more than high peak data rates. Sure enough, Wi-Fi~7 is projected to support at least 30~Gbps per AP, about four times as fast as Wi-Fi~6, while ensuring backward compatibility and coexistence with legacy devices in the 2.4, 5, and 6~GHz unlicensed bands. However, the 802.11be Task Group (TG) also recognised the need for---and is aiming at---lower latencies and higher reliabilities to enable time-sensitive networking (TSN) use cases \cite{AdaCarBel2019}. The former is seen as an enabler for real-time applications including augmented and virtual reality, gaming, and cloud computing, demanding delay times reduced to below 5~ms. The latter is critical for next-generation factories and enterprises, where Wi-Fi may need to guarantee three or more `nines' of reliability to aim at replacing some wired communications. 

To speed up the development and commercialisation of Wi-Fi 7, 
whose timeline is illustrated in Fig. \ref{fig:timeline}, 
the 802.11be TG deviated from the conventional single-phase development cycle, and agreed on two phases.
The first one will place the spotlight on a set of features deemed of high-priority according to their gain/complexity ratio, time for standardization and implementation, as well as associated interest and market needs. More on this in the next section.

 
\section{A Primer on the Future Wi-Fi 7 \label{sec:features}}

At the time of writing, the 802.11be TG is actively defining the basic 
functional operations that will be included in the standard. These are collected in the Specification Framework Document (SFD), from which the consecutive drafts of the standard will be derived \cite{2020SFD80211be}. We focus on digesting its latest updates in the sequel.

\subsection{Release 1 Features \label{sec:release1Features}}
As indicated in Fig. \ref{fig:timeline}, 
Release 1 (R1) features are foreseen to reach a mature specification in Draft 1.0, 
due in May 2021, 
with the possibility to further expand and refine them until the release of Draft 2.0 in March 2022.
They include: 

\subsubsection{Multi-Link Operation}

802.11be targets efficient operations in all the available bands, i.e. 2.4, 5, and 6 GHz, for load balancing, multi-band aggregation, and simultaneous downlink/uplink transmission \cite{LopGarGal2019}. In 802.11be, a multi-link device (MLD) is defined as one with multiple affiliated APs or stations (STAs), and a single MAC service access point (SAP) to the above logical link control (LLC) layer. A MAC address that uniquely identifies the MLD management entity is also introduced. The most relevant features for the control and operation of multiple links are summarized as follows:

\begin{itemize}
    \item \emph{Multi-link discovery and setup:} MLDs will be capable of dynamically updating their ability for simultaneous frame exchange on each pair of links. Moreover, each individual AP/STA may also provide information on the operational parameters of the other affiliated APs/STAs within the same MLD.

    \item \emph{Traffic-link mapping:} Upon multi-link setup, all traffic identifiers (TIDs)---used to classify frames based on their quality of service (QoS)---are mapped to all setup links. An update of this mapping can be subsequently negotiated by any MLD involved. In addition, the recipient MLD will utilize a single reordering buffer for QoS data frames of the same TID transmitted over multiple links.

    \item \emph{Channel access and power saving:} Each AP/STA of a MLD performs independent channel access over its links and maintains its own power state. 
    To facilitate an efficient STA power management, APs may also utilize an enabled link to carry indications of buffered data for transmission on other links.
\end{itemize}

\subsubsection{Low-Complexity AP Coordination}


802.11be will support multi-AP coordination, with APs advertising their capabilities in beacons / management frames. Coordinated spatial reuse (CSR) is one low-complexity implementation that may be included in R1. In CSR, a \emph{sharing} AP that has acquired a transmission opportunity (TXOP) can trigger one or more other \emph{shared} APs to perform simultaneous transmission with appropriate power control and link adaptation. This coordination will create more spatial reuse opportunities and reduce the number of collisions when compared with the spatial reuse schemes available in 802.11ax.

\subsubsection{Direct Enhancements of 802.11ax}

The 802.11be TG will also specify a number of  direct upgrades to the current 802.11ax standard. These include:
\begin{itemize}
    \item 
    \emph{Support of 320 MHz transmissions}---doubling the 160 MHz of 802.11ax.
    \item
    \emph{Use of higher modulation orders}, optionally supporting 4096-QAM---up from 1024-QAM in 802.11ax---with a strict -38 dB requirement on the error vector magnitude (EVM) at the transmitter.
    \item    
    \emph{Allocation of multiple resource units}, i.e. groups of OFMDA tones, per STA. This extra degree of flexibility leads to more efficient spectrum utilization \cite{KhoLevAky2020}.
\end{itemize}

\subsection{Release 2 Features \label{sec:release2Features}}
Although R2 features will be formalized in Draft 3.0 and Draft 4.0,
due respectively in Nov. 2022 and Nov. 2023, the 802.11be TG has already started its work on them and has achieved a notable progress within the SFD. The main features are detailed below:

\subsubsection{MIMO Enhancements}

802.11be will double the maximum number of supported single-user MIMO (SU-MIMO) and multi-user MIMO (MU-MIMO) spatial streams to 16, with a consequent increase in capacity. In the case of MU-MIMO, the 802.11be TG agreed on limiting the maximum number of spatially multiplexed STAs and spatial streams per STA to 8 and 4, respectively. The aforesaid limits help controlling both the MIMO precoder complexity and channel state information (CSI) acquisition overhead. To further curb this overhead, implicit CSI sounding is being investigated as an optional mode.

\subsubsection{Hybrid Automatic Repeat Request (HARQ)}
R2 will likely see the introduction of HARQ, where devices do not discard erroneous information but attempt to softly combine it with retransmitted units to increase the probability of correct decoding. While the SFD does not include any HARQ-related procedures at the time of writing this article, the 802.11be TG has examined different HARQ units---MAC protocol data units (MPDUs) or PHY codewords---and extensively evaluated the performance/complexity trade-offs.



\subsubsection{Low-Latency Operations}
Given the business appeal of TSN \cite{AdaCarBel2019}, the SFD will also collect the protocol enhancements dedicated to reduce worst-case latency and make a leap in reliability. Conceivably, such solutions may rely on multi-link operations---providing a differentiated QoS per link---, or on AP coordination, for a more aggressive spectrum reuse and fewer harmful collisions.

\subsubsection{Advanced AP Coordination}
To reach the full 
potential of multi-AP coordination, the 802.11be TG has agreed to support the following three schemes:
\begin{itemize}
    \item 
    \emph{Coordinated OFDMA:} 
    In 802.11be, an AP that obtains a TXOP will be able to share its frequency resources in multiples of 20 MHz channels with a set of neighbouring APs. 
    For the sake of efficiency, the sharing AP may request neighbouring APs to report their resource needs.
    

    


    \item
    \emph{Joint single- and multi-user transmissions:} 
    Collectively sending data to their connected STAs requires APs to bound their phase synchronization errors and timing offsets. Joint transmissions have been found to provide gains when considering reasonable values for these impairments, provided that an adequate backhauling is available. Since collaborative APs require CSI from both associated and non-associated STAs, 802.11be will define a joint multi-AP sounding scheme. This way, APs will simultaneously transmit their sounding frames and the addressed STAs will convey CSI feedback pertaining to all APs.

    \item
    \emph{Coordinated beamforming:} 
    This technique exploits the capability of modern multi-antenna APs to spatially multiplex their STAs, while jointly placing radiation nulls to/from neighbouring non-associated STAs. While the CSI required to steer radiation nulls can be obtained via the aforementioned joint multi-AP sounding scheme, CBF can also take advantage of the simpler sequential sounding procedure that will be part of 802.11be. Moreover, CBF does not require joint data processing as each STA transmits/receives data to/from a single AP, therefore significantly diminishing backhauling needs w.r.t. joint transmissions. Since CBF can deliver substantial throughput and latency enhancements while keeping complexity at bay, we further explore it in the next section.
\end{itemize}


\begin{figure*}[t!]
\begin{centering}
	\includegraphics[width=16cm]{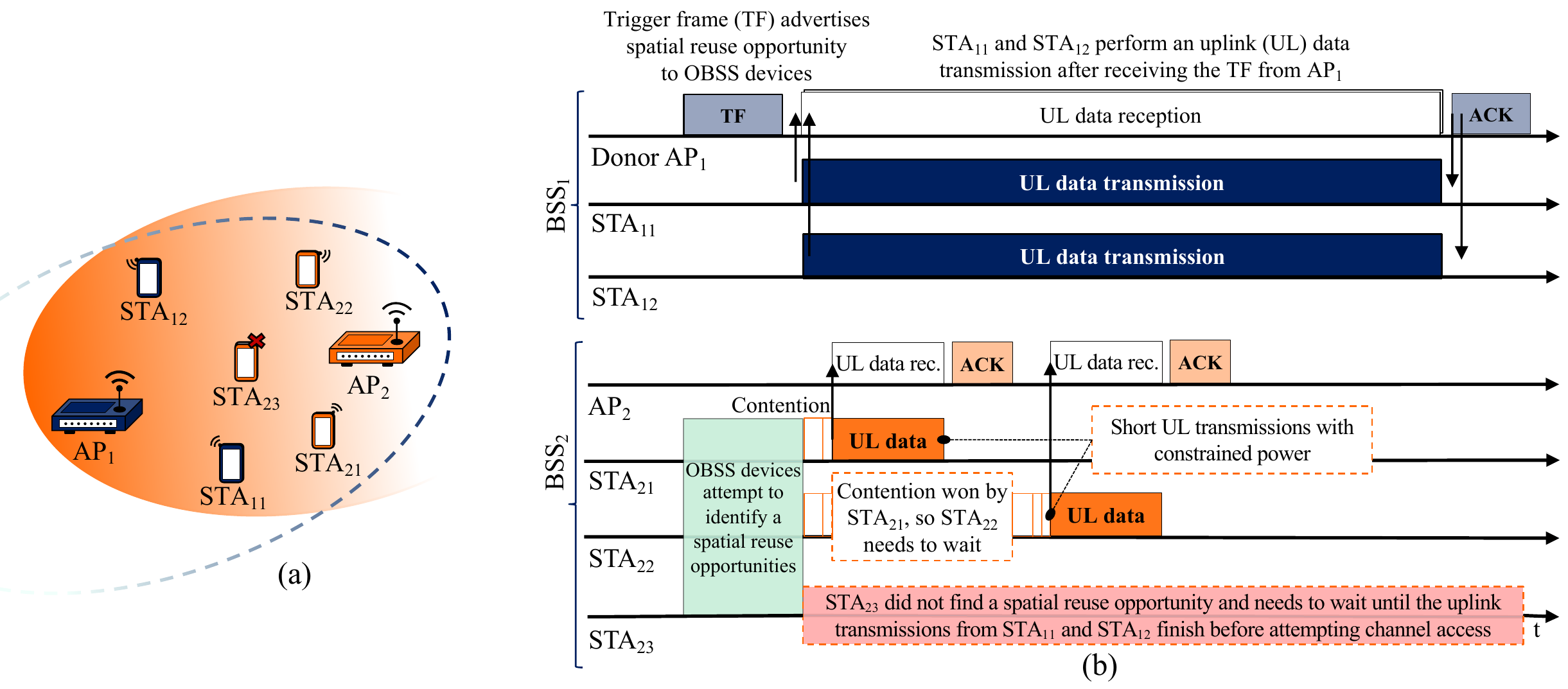} 
	\caption{Illustration of the PSR framework.}
\label{fig:psrFramework}
\end{centering}
\end{figure*}

\section{Augmenting Spatial reuse via Multi-AP Coordinated Beamforming} 
\label{sec:coordinatedBeamforming}


A certain consensus has been reached on the fact that reliability and low-latency features may potentially build upon 802.11ax, thus facilitating backward compatibility, product certification and market adoption.
To this end, \emph{parameterised spatial reuse} (PSR) in 802.11ax is an appealing building block, as it allows for a dynamic cooperation between devices of different BSSs \cite{Wilhelmi2019}.
In the following, we introduce the PSR framework, discuss its shortcomings, and explain how it can be extended through multi-AP coordination to curb latency and increase reliability in 802.11be.


\subsection{Parameterised Spatial Reuse in 802.11ax} 
\label{sec:spatialReuse80211ax}

In PSR, a donor AP intending to perform uplink reception may opportunistically provide TXOPs to an overlapping BSSs (OBSSs) through a trigger frame. In its basic form, a trigger frame can be seen as a scheduling grant, providing information and timing for the subsequent uplink transmissions. When enabling PSR, a donor AP leverages a trigger frame to invite OBSSs devices to reuse the spectrum concurrently with its uplink reception, provided that they meet certain interference conditions.


To provide a more detailed description of the PSR framework, let us consider the example of \figurename~\ref{fig:psrFramework}(a) with two BSSs, where:
\begin{itemize}
\item
${\rm BSS}_1$ is comprised of ${\rm AP}_1$, ${\rm STA}_{11}$, and ${\rm STA}_{12}$; 
whereas 
\item
${\rm BSS}_2$ includes ${\rm AP}_2$, ${\rm STA}_{21}$, ${\rm STA}_{22}$, and ${\rm STA}_{23}$.
\end{itemize}

\figurename~\ref{fig:psrFramework}(b) illustrates how ${\rm AP}_1$, after getting channel access, 
starts the PSR process by transmitting a trigger frame. This trigger frame has a dual functionality:
\begin{itemize}
\item
Conveying the synchronisation and scheduling information necessary for the uplink transmissions of its associated ${\rm STA}_{11}$ and ${\rm STA}_{12}$; 
and 
\item
Advertising a spatial reuse opportunity to the OBSS devices, 
with such opportunity spanning the subsequent uplink data reception of ${\rm AP}_1$.
\end{itemize}
To guarantee that transmissions taking advantage of the spatial reuse opportunity do not impact the uplink data reception  of ${\rm AP}_1$, the trigger frame includes a \emph{PSR field}. This field contains information about 
\emph{i)} the maximum interference level that ${\rm AP}_1$ can receive without harming its uplink reception,
and \emph{ii)} the transmit power of ${\rm AP}_1$, to facilitate interference calculations. 
Upon reception of the trigger frame, OBSS devices measure its received power level and, based on the information provided in the PSR field, determine whether they can access the medium and with what transmit power.

\begin{figure*}[t!]
\begin{centering}
	\includegraphics[width=17.5cm]{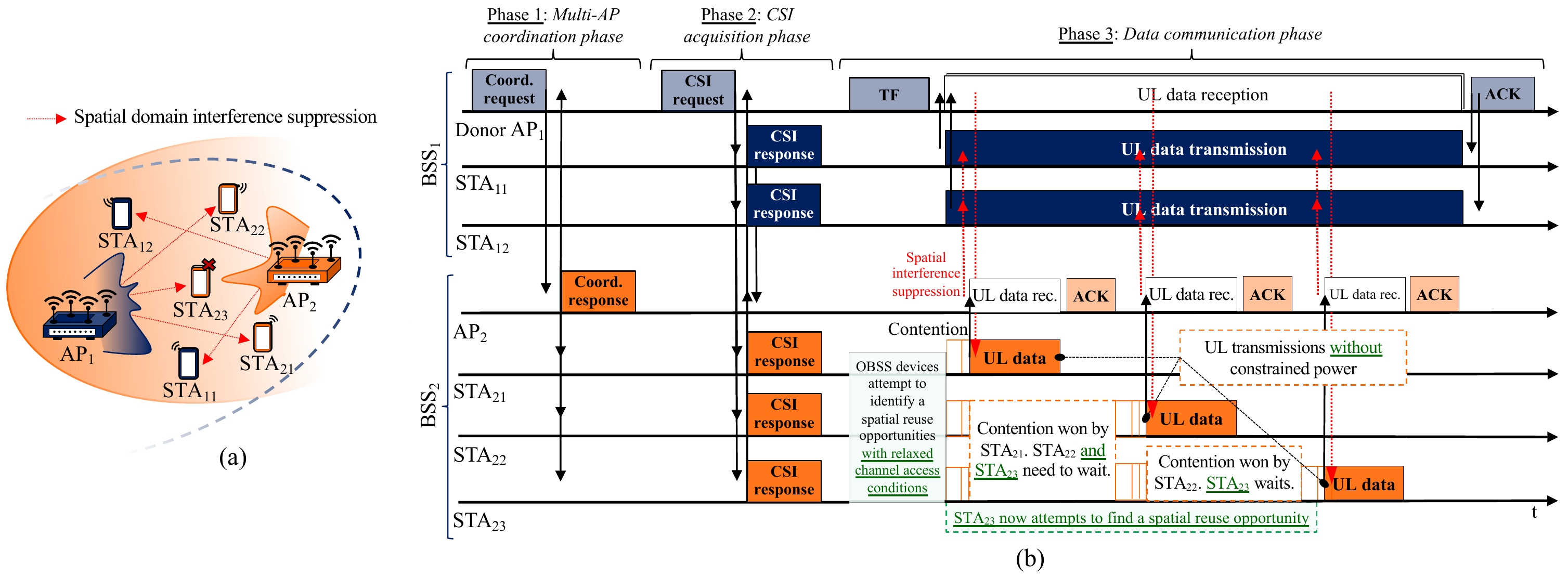} 
	\caption{Illustration of the proposed coordinated beamforming protocol.}
\label{fig:psrFrameworkCoordinated}
\end{centering}
\end{figure*}

In our example of \figurename~\ref{fig:psrFramework}(b), ${\rm STA}_{21}$, ${\rm STA}_{22}$ and ${\rm STA}_{23}$ all have uplink data to transmit. However, only ${\rm STA}_{21}$ and ${\rm STA}_{22}$ independently determine that they can contend for the medium. Unfortunately, ${\rm STA}_{23}$ cannot contend for channel access, since its proximity to ${\rm AP}_1$ prevents it from meeting the interference conditions set by the latter. As a result, ${\rm STA}_{21}$ accesses the channel first to send its short packet, making sure that the corresponding acknowledgement (ACK) frame is received within the duration of the uplink transmissions triggered by ${\rm AP}_1$. As long as such duration allows, ${\rm STA}_{22}$ will also have the chance to re-contend for the channel and transmit.


\subsubsection{Benefits of PSR}
\label{sec:nbenefitsSpatialReuse80211ax}

Overall, thanks to the PRS framework, APs and STAs can gain channel access while traditionally they would have not. This improves spatial reuse, and in turn:
\begin{itemize}
\item
Increases the network throughput, 
as it allows more concurrent transmissions;
\item
Increases the STA file throughput, 
as STAs spend less time in contention; 
and importantly,
\item
Reduces latency as STAs with time-sensitive short-file traffic may not need to wait until broadband STAs terminate their lengthy transmissions. This is the case of $\text{STA}_{21}$ and $\text{STA}_{22}$ in \figurename~\ref{fig:psrFramework}.
\end{itemize}

\subsubsection{Challenges of PSR} 
\label{sec:issuesSpatialReuse80211ax}

While the PSR framework allows for a larger spatial reuse, 
two fundamental challenges have been identified within the 802.11be forum:
\begin{itemize}
\item 
Devices taking advantage of a spatial reuse opportunity must lower their transmit power to limit the interference generated. In some cases, as for $\text{STA}_{21}$ and $\text{STA}_{22}$ in \figurename~\ref{fig:psrFramework}, this translates into a reduced throughput. In other cases, as for $\text{STA}_{23}$, devices cannot even access spatial reuse opportunities as their maximum allowed transmit power is insufficient to reach their receiver. 
\item 
Devices taking advantage of a spatial reuse opportunity are unaware---and have no control over---the interference perceived by their respective receivers. In \figurename~\ref{fig:psrFramework}, this entails that uplink transmissions from $\text{STA}_{21}$ and $\text{STA}_{22}$ to $\text{AP}_2$ are likely to fail if $\text{STA}_{11}$ and $\text{STA}_{12}$ are near $\text{AP}_2$, since $\text{AP}_2$ will receive a non-negligible amount of interference from $\text{STA}_{11}$ and $\text{STA}_{12}$. 
\end{itemize}

The two above shortcomings hamper the effectiveness of the existing PSR framework in a variety of setups. These include high-density scenarios, or those where devices deal with latency-sensitive data traffic and cannot afford transmission failures or excessive channel access waiting times. 


\subsection{Coordinated Beamforming in 802.11be} 
\label{sec:spatialReuse80211be}


802.11be aims at taking existing spatial reuse capabilities to a whole new level through CBF, i.e. by letting collaborative APs suppress incoming OBSS interference in the spatial domain. 
Recent experimental studies have demonstrated that, when compared to a single-antenna system, a four-antenna AP serving one STA is capable of suppressing up to 10 dB of interference towards a neighbouring link~\cite{bertizzolo2020cobeam}.
Motivated by these promising results, we now detail an illustrative protocol that implements CBF by smoothly building on the PSR framework. 

Let us consider the uplink transmission scenario of \figurename~\ref{fig:psrFrameworkCoordinated}(a). 
This setup resembles that of \figurename~\ref{fig:psrFramework}(a),
but $\text{AP}_1$ and $\text{AP}_2$ are now equipped with eight antennas. The three phases of the proposed CBF protocol---the first two being common to the CBF and joint transmission implementations discussed so far in 802.11be---are illustrated in \figurename~\ref{fig:psrFrameworkCoordinated}(b) and described as follows.
\subsubsection{Multi-AP Coordination}
During this phase, 
the two or more collaborative APs exchange control frames with a twofold purpose:
	\begin{itemize}
	\item 
	\emph{Coordination set establishment and maintenance:}  
	For CBF to be effective, APs need to communicate with OBSS STAs, e.g. for acquiring the necessary CSI to place radiation nulls on a specific spatial location. 
	To this end, an inter-BSS coordination set is defined between the collaborative APs, which must contain the IDs of all APs and STAs participating to the CBF transmission. These IDs may be kept in memory by all devices involved, which will not discard---as they traditionally would---the relevant frames generated by OBSS devices included in their coordination sets. Once defined, an inter-BSS coordination set may be updated in a semi-static manner, i.e. after tens or hundreds of TXOPs. 
	\item 
	\emph{Dynamic coordination of the subsequent spatial reuse opportunity:} 
	Once the donor $\text{AP}_1$ obtains a TXOP, it needs to advertise the incoming uplink-triggered transmission and, together with the devices in its coordination set, determine which STAs will be involved in the subsequent CSI acquisition and data communication phases. 
	In the example of \figurename~\ref{fig:psrFrameworkCoordinated}(b), 
	$\text{AP}_2$ replies to the dynamic coordination frame sent by $\text{AP}_1$, 
	indicating which of its STAs would most benefit from being granted a safe spatial reuse opportunity, 
	e.g. $\text{STA}_{21}$ and $\text{STA}_{22}$ as they typically generate latency-sensitive traffic.
	\end{itemize}
\subsubsection{CSI Acquisition}
During this phase, thanks to the previous coordination, both $\text{AP}_1$ and $\text{AP}_2$ acquire CSI from relevant intra-BSS and OBSS devices only. Such CSI is necessary in order to design a filter for spatial multiplexing and bidirectional interference suppression in the subsequent communication phase. 
Importantly, as OBSS devices are being addressed for CSI acquisition, they become aware that an OBSS AP will shortly offer them a spatial reuse opportunity with more favourable channel access conditions. As there is no need for new specific signalling to trigger data communications, the 802.11ax trigger frame can be used for this purpose. This bring obvious benefits to legacy STAs,
which may keep applying the legacy PSR framework of 802.11ax in a seamless manner.
\subsubsection{Data communication}
The implementation of the two previous phases addresses the two fundamental challenges of the 802.11ax PSR framework highlighted in the previous section, making spatial reuse transmissions from $\text{STA}_{21}$, $\text{STA}_{22}$ and  $\text{STA}_{23}$ much more likely to timely succeed in adverse conditions. This is because:
\begin{itemize}
\item 
$\text{STA}_{21}$, $\text{STA}_{22}$ and even $\text{STA}_{23}$ are more likely to find spatial reuse opportunities and use their maximum transmission power. This is thanks to the spatial interference suppression performed by $\text{AP}_1$, which facilitates the advertisement of a relaxed the channel access conditions for the relevant OBSS devices.
\item 
$\text{AP}_2$ is now capable of suppressing the incoming interference generated by $\text{STA}_{11}$ and $\text{STA}_{12}$, 
while receiving the uplink transmissions from $\text{STA}_{21}$, $\text{STA}_{22}$ and $\text{STA}_{23}$.
\end{itemize}

\section{Performance of 802.11be Coordinated Beamforming}
\label{sec:performanceEvaluation}

\begin{table}
\centering
\caption{System-level simulation parameters}
\vspace{-0.15cm}
\label{table:parameters}
\def\arraystretch{1.1}
\begin{tabulary}{\columnwidth}{ |p{3.4cm} | p{4.5cm} | }
\hline
\rowcolor{LightBlue}
\textbf{Parameter} 		& \textbf{Description} 						\\ \hline
\rowcolor{LightCyan}
\textbf{Deployment} 		&  							\\ \hline
AP 2D locations 		&  $(10\text{m}, 10\text{m})$ and $(25\text{m}, 10\text{m})$ 				\\ \hline
STA 2D distribution / height		& Uniform deployment / $h = 1\text{m}$					\\ \hline
AP-STA association criterion 		& Strongest average received signal
\\ \hline
\rowcolor{LightCyan}
\textbf{Traffic model} 		&  							\\ \hline
Broadband STAs 		& FTP 3 with 100\,Mbits/s of offered traffic and a packet size of 0.5 MBytes \cite{3GPP38824} 		\\ \hline
Augmented reality STAs		& Constant arrival rate at 10 ms frequency and 32 bytes packet size \cite{3GPP38824} 	\\ \hline
\rowcolor{LightCyan}
\textbf{Channel model} 		&  							\\ \hline
Spatial channel model 		& 3GPP 3D InH for all links \cite{3GPP38901} 					\\ \hline
Thermal noise 			& -174 dBm/Hz spectral density						\\ \hline
\rowcolor{LightCyan}
\textbf{MAC} 			&  							\\ \hline
Maximum TXOP length 		& 4 ms 							\\ \hline
MCS selection algorithm		& SINR-based selection 						\\ \hline
AP scheduling policy 		& Spatial multiplexing of as many STAs as possible of the same traffic class 			\\ \hline
\rowcolor{LightCyan}
\textbf{PHY} 			&  							\\ \hline
Carrier frequency / bandwidth		& 5.18 GHz / 80 MHz						\\ \hline
AP/STA maximum TX power		& $P_{\textrm{max}} = 24/15$ dBm					\\ \hline
AP/STA antennas 		& $4\times 2$ omni array / 1 omni					\\ \hline
AP receive spatial filter 		& ZF with up to 4 inter-BSS nulls \cite{8121870}				\\ \hline
AP interference suppression 		& Imperfect, 10 dB per device \cite{bertizzolo2020cobeam, 8121870}			\\ \hline
AP/STA noise figure 		& $F_{\textrm{dB}} = 7/9$\,dB						\\ \hline	 
\end{tabulary}
\vspace{-0.5cm}
\end{table}

We now quantify the latency enhancements provided by the CBF scheme described throughout the previous section. With this objective, we consider a deployment with $2$ ceiling-mounted APs, each equipped with $8$ antennas, and $24$ STAs uniformly distributed across an indoor enterprise scenario of $35$m$\times20$m$\times3$m. Out of those $24$ STAs, $16$ STAs generate uplink broadband traffic, and the remaining 8 STAs generate uplink latency-sensitive augmented reality traffic. 
Since our main objective is to guarantee on-time delivery of the augmented reality traffic, 
APs granting spatial reuse opportunities will suppress interference from neighbouring augmented reality STAs generating the strongest interference---which usually correspond to those located closest.
The results in this section are the outcome of intricate standard-contributed system-level simulations, and Table~\ref{table:parameters} provides the basic setup that suffices for their understanding. Interested readers may find the full set of simulation parameters in \cite{2020PerformancePSRWithNullSteering}.

\begin{figure}[t!]
\begin{centering}
	\includegraphics[width=0.48\textwidth]{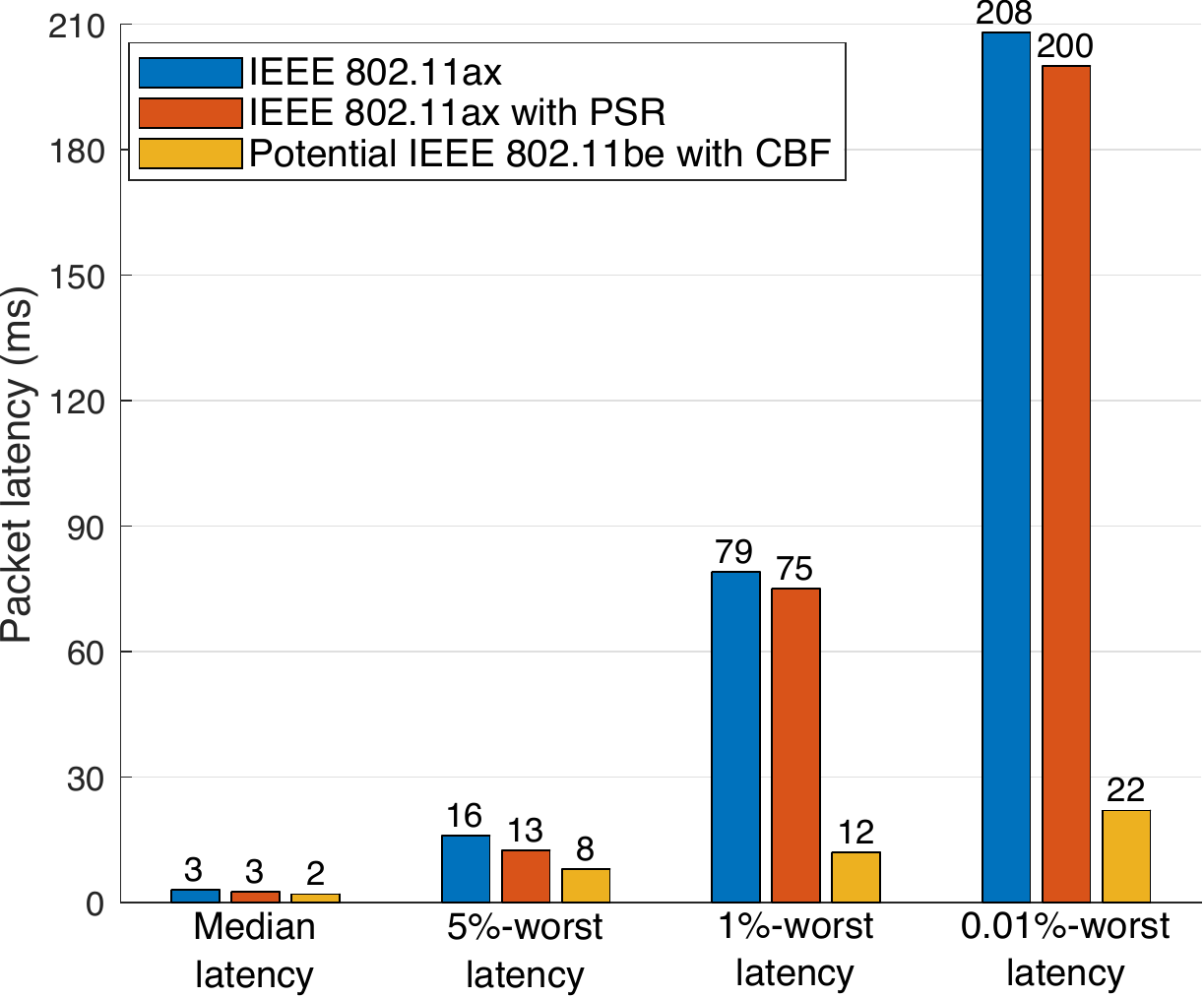} 
	\caption{Median and worst-case latencies (ms) experienced by the augmented reality STAs. 
	Three systems are evaluated: 
	\emph{1)} IEEE 802.11ax without spatial reuse, 
	\emph{2)} IEEE 802.11ax with PSR capabilities, and 
	\emph{3)} a IEEE 802.11be system with CBF capabilities.}
\label{fig:latencyResults}
\end{centering}
\end{figure}

\figurename~\ref{fig:latencyResults} represents the median, $5$\%-, $1$\%-, and $0.01$\%-worst MAC layer latencies experienced by the augmented reality STAs for three different setups:
\begin{itemize}
\item 
\emph{A setup where IEEE 802.11ax devices do not have spatial reuse capabilities:} 
The results of \figurename~\ref{fig:latencyResults} support the broad agreement that IEEE 802.11-based systems may be capable of delivering low latency, 
but struggle to maintain a consistent performance in the worst cases \cite{AdaCarBel2019}. 
Indeed, we can observe that latency remains below $3$\,ms around $50\%$ of the time in the scenario considered, but it dramatically grows above $200$\,ms for the $0.01\%$ worst-cases.
This is mostly due to the combined impact of the random channel access mechanism as well as the collisions that lead to retransmissions.
\item 
\emph{A setup with PSR-capable IEEE 802.11ax devices:} 
\figurename~\ref{fig:latencyResults} illustrates that the implementation of PSR does not help in substantially reducing the worst-case delays. 
This is because, 
similarly to what occurs to $\text{STA}_{23}$ in \figurename~\ref{fig:psrFramework}, 
augmented reality STAs are not sufficiently far apart from their neighbouring APs in the dense scenario considered. 
This prevents these latency-sensitive STAs from finding spatial reuse opportunities, 
since the channel access conditions they need to adhere to prevent harmful interference are too stringent, as detailed in Sec. \ref{sec:issuesSpatialReuse80211ax}. 
\item 
\emph{A setup where devices implement the IEEE 802.11be CBF scheme described in the previous section:} 
Importantly, 
the results of \figurename~\ref{fig:latencyResults} illustrate how the proposed scheme provides a substantial reduction of the worst-case latencies when compared to the other IEEE 802.11ax systems. 
Indeed, we can observe that the system with multi-AP coordination capabilities drives down the $0.01$\%-worst-case latencies by a factor of $\approx9\times$ with respect to a PSR-capable system. 
This significant performance enhancement is a direct consequence of
\emph{i)} the substantially larger number of spatial reuse opportunities found by augmented reality STAs, due to their relaxed channel access conditions,
and \emph{ii)} the OBSS interference mitigation provided in the spatial domain, 
which maximises the chances of performing successful data transmissions.
\end{itemize}
Although not shown for brevity, 
it should be remarked that the throughput of the broadband STAs is approximately maintained for the three systems under evaluation~\cite{2020PerformancePSRWithNullSteering}.


\section{Conclusion\label{sec:Conclusion}}
Next-generation Wi-Fi will unlock access to gigabit, reliable and low-latency communications, reinventing manufacturing and social interaction through digital augmentation. In this article, we detailed the steps taken by IEEE 802.11be towards Wi-Fi 7, the latest agreements on its technical features and its recently updated timeline. We put forward the importance of spatial reuse through multi-AP coordinated beamforming, sharing implementation details and standard-compliant simulations to illustrate its benefits. Looking ahead, further research is needed to blend such techniques into time-sensitive networking protocols, with the overarching goal of making wireless the new wired for both our homes and industries. 




\bibliographystyle{IEEEtran}
\bibliography{Ming_library}

%
%
%
%
%

\end{document}